\documentclass[12pt]{iopart}
\usepackage{iopams}
\usepackage{amsthm}
\usepackage{graphicx}
\usepackage{mypackage}

\def\varstackrel#1#2{\mathrel{\mathop{#2}\limits_{#1}}}

\theoremstyle{plain}
\newtheorem{theorem}{Theorem}
\newtheorem{proposition}[theorem]{Proposition}
\newtheorem{lemma}[theorem]{Lemma}

\newtheorem{claim}[theorem]{Claim}

\theoremstyle{definition}
\newtheorem{definition}{Definition}

\theoremstyle{remark}

\newcommand{\epimmP}{e^{2\pi{}\rmi\frac{mm'}{P}}}
\newcommand{\piP}{\frac{2\pi{}\rmi}{P}}

\newcommand{\summ}{\sum_{m=0}^{P-1}}
\newcommand{\summp}{\sum_{m'=0}^{P-1}}

\newcommand{\sinL}[1]{\frac{\sin[\pi #1]}{2^{p-j-1}\sin[\pi #1 /2^{p-j-1}]}}
\newcommand{\cosmj}[1]{\cos\left[\pi #1 /2^{p-j}\right]}
\newcommand{\sinmj}[1]{\sin\left[\pi #1 /2^{p-j}\right]}
\newcommand{\sinM}[1]{\frac{2^{p-j}\sin[\pi #1 /2^{p-j}]}{2^p\sin[\pi #1 /2^p]}}

\newcommand{\sinNoP}[1]{\frac{\sin[\pi #1]}{P\sin[\pi #1 /P]}}

\newcommand{\fourGroverstate}[1]
{
 \cef{u}{}{#1} \ket{b} + \cef{v}{}{#1} \ket{g}
 + \cef{u}{e}{#1} \ketebasis{b}{#1}
 + \cef{v}{e}{#1} \ketebasis{g}{#1}
}

\newcommand{\rhalf}{\left( r+\frac{1}{2} \right)}
\newcommand{\eirt}{e^{\rmi{}r\theta}}
\newcommand{\emirt}{e^{-\rmi{}r\theta}}
\newcommand{\eimrt}{e^{\rmi(m-r)\theta}}
\newcommand{\emimrt}{e^{-\rmi(m-r)\theta}}
\newcommand{\epimr}{e^{\pi{}\rmi(m-r)}}
\newcommand{\epifP}{e^{ \frac{\pi{}\rmi{}f}{P} }}
\newcommand{\empifP}{e^{ -\frac{\pi{}\rmi{}f}{P} }}

\newcommand{\cef}[3]{{#1}_{#2}^{(#3)}}

\newcommand{\ub}{\cef{u}{}{b}}
\newcommand{\vb}{\cef{v}{}{b}}
\newcommand{\ueb}{\cef{u}{e}{b}}
\newcommand{\veb}{\cef{v}{e}{b}}

\newcommand{\ug}{\cef{u}{}{g}}
\newcommand{\vg}{\cef{v}{}{g}}
\newcommand{\ueg}{\cef{u}{e}{g}}
\newcommand{\veg}{\cef{v}{e}{g}}

\newcommand{\up}{\cef{u}{}{\phi}}
\newcommand{\vp}{\cef{v}{}{\phi}}
\newcommand{\uep}{\cef{u}{e}{\phi}}
\newcommand{\vep}{\cef{v}{e}{\phi}}

\newcommand{\ebasis}[2]{e_{#1}^{(#2)}}
\newcommand{\ketebasis}[2]{\ketfix{\ebasis{#1}{#2}}}

\newcommand{\ebb}{\ebasis{b}{b}}
\newcommand{\egb}{\ebasis{g}{b}}
\newcommand{\ebg}{\ebasis{b}{g}}
\newcommand{\egg}{\ebasis{g}{g}}
\newcommand{\ebp}{\ebasis{b}{\phi}}
\newcommand{\egp}{\ebasis{g}{\phi}}

\newcommand{\ketebb}{\ketebasis{b}{b}}
\newcommand{\ketegb}{\ketebasis{g}{b}}

\newcommand{\ketebg}{\ketebasis{b}{g}}
\newcommand{\ketegg}{\ketebasis{g}{g}}

\newcommand{\ketebp}{\ketebasis{b}{\phi}}
\newcommand{\ketegp}{\ketebasis{g}{\phi}}

\newcommand{\ceffirst}[2]{c_{{#2}}^{(#1)}}

\newcommand{\dasc}[1]{d_{\mbox{asc}}^{(#1)}}
\newcommand{\ddes}[1]{d_{\mbox{des}}^{(#1)}}

\newcommand{\keteb}{\ket{e_b}}
\newcommand{\keteg}{\ket{e_g}}

\begin{document}

\title
{Theoretical Analyses of Quantum Counting against Decoherence Errors}

\author{Jun Hasegawa\dag\ddag\ and Fumitaka Yura\ddag}

\address{\dag\ Department of Computer Science,
Graduate School of Information Science and Technology,
the University of Tokyo.
7-3-1 Hongo, Bunkyo-ku, Tokyo 113-0033, Japan.}

\address{\ddag\ ERATO Quantum Computation and Information Project, JST.
Hongo White Building, 5-28-3 Hongo, Bunkyo-ku, Tokyo 113-0033, Japan.}

\eads{\mailto{hasepyon@is.s.u-tokyo.ac.jp}, \mailto{yura@qci.jst.go.jp}}


\begin{abstract}
 In this paper, we analyze the quantum counting under the decoherence,
 which can find the number of solutions
 satisfying a given oracle.
 We investigate probability distributions
 related to the first order term of the error rate
 on the quantum counting with the depolarizing channel.
 We also implement two circuits for the quantum counting
 -- the {\itshape ascending-order} circuit and
 the {\itshape descending-order} circuit --
 by reversing ordering of application of controlled-Grover operations.
 By theoretical and numerical calculations for probability distributions,
 we reveal the difference of probability distributions
 on two circuits
 in the presence of decoherence
 and show that the ascending-order circuit
 is more robust against the decoherence
 than the descending-order circuit.
 This property of the robustness is applicable to
 the phase estimation such as the factoring.
\end{abstract}

\maketitle


\section{Introduction}

Since Grover demonstrated the database search algorithm~\cite{Grover96grover},
many applications of this algorithm have been
studied~\cite{DH96minimum,Grover96median}.
The {\itshape quantum counting}~\cite{BHT98counting} is 
one of the most important applications
with using the quantum Fourier transform.
Given a quantum oracle,
the quantum counting can find the number $t$ of solutions
of the oracle through in $N$ elements with
$\Or(\sqrt{tN})$ operations,
whereas $\Or(N)$ operations are required on a classical computer.
The quantum counting is considered to be used
for \textbf{NP}-complete problems
because the quantum counting helps us
determine the existence of a solution of these problems
quadratically faster than a classical algorithm,
solving whether the number of solutions
is zero or non-zero.
Recently, there have been works
for numerical integrals based on the quantum counting,
called quantum summation~\cite{AW99integrals,TW02path,Heinrich02summation}.
One can estimate the value of integrals with $\Or(1/\varepsilon)$ operations
for the desired accuracy $\varepsilon$,
while $\Or(1/\varepsilon^2)$ operations are needed
by classical Monte Carlo method.

When we realize such quantum algorithms,
decoherence is inevitable since our apparatus are
surrounded with environment and open systems for us.
Therefore treating the decoherence should be always
one of significant problems.
%
%
In the previous works, there have been
only a few analyses of the decoherence,
related to especially Shor's factoring~\cite{Shor94factorization}
and Grover's database search algorithm.
Azuma~\cite{azuma02decoherence}
investigated the decoherence on Grover's algorithm
by calculating quantum states in detail
up to the fifth order term of the decoherence
for $\sigma_z$ errors.
Shapira et al.~\cite{SMB03noise}
dealt with the decoherence
on Grover's algorithm
by changing the Hadamard transformation
into some distorted operation.
Yu et al.{}~\cite{YS99SU}
represented disturbed unitary operations
by the decoherence
by focusing on quantum Hamiltonian
and 
analyzed the evolution of quantum system
on Grover's database search algorithm.
Sun et al.{}~\cite{SZL98decoherence}
showed effects of environment
based on the dynamic approach
for quantum measurement
through the example of the factoring.
Several researchers have been investigated the decoherence
by numerical calculations.
%
Obenland et al.{}~\cite{OD99simulating}
simulated the circuits which factored
the numbers 15, 21, 35, and 57
as well as circuits that solved
the database search
for a trapped ion quantum computer.
Niwa et al.{}\
implemented
the general-purpose parallel simulator for
quantum computing
which could simulate
not only the factoring and the database search
algorithm~\cite{NMI02simulator}
but also quantum error correcting codes~\cite{Niwa02ErrorCorrect},
and revealed influences of the decoherence
and another quantum error -- operational error --
to these algorithms.
Although the decoherence on the factoring and the database search
have been analyzed in a variety of ways,
there have been no analysis of the decoherence
on the quantum counting.


In this paper, we investigate the decoherence
related to the first order term of error rate
on the quantum counting
and expand our results to the phase estimation algorithm
such as the factoring.
The quantum counting is composed of two registers,
called the {\it first register} and the {\it second register}.
We assume the depolarizing channel as error models
and calculate probability distributions on the quantum counting
in two cases that the decoherence error occurs on each register.

We have another purpose in this paper to reveal
which implementation for the quantum counting
and one of the key quantum algorithms, the {\it phase estimation},
is robust against decoherence.
Two quantum algorithms estimate a phase of
a unitary operator by using the quantum Fourier transform
and can be implemented in many ways
by changing ordering of application of the unitary operations.
In order to show the difference of effects of decoherence
among different implementations,
we implement two typical circuits --
the {\it ascending-order} circuit and the {\it descending-order}
circuit
and compare probability distributions on these circuits
in the presence of decoherence.

In the case that the depolarizing channel is applied
on the first register in the quantum counting,
we show that a probability distribution has
many peaks caused by the decoherence
at a distance of the power of two from correct peaks.
We also show that 
probability distributions between the ascending-order
and the descending-order circuit are the same.
On the other hand, in the case of the second register,
we reveal probability distributions on two circuits
are completely different.
In the ascending-order case,
wrong outputs near correct one are obtained
by the quantum counting,
while in the descending-order case,
two wrong outputs $0$ and $N$ 
are obtained with high probability
independently of an input oracle.
Additionally,
the correct output is obtained
with higher probability
on the ascending-order circuit
than on the descending-order one.
It follows that the ascending-order implementation for the quantum counting
is more robust against the decoherence.

We also show robustness against decoherence on the phase estimation
with the ascending-order.
Finally, we discuss weakness against decoherence on an efficient
implementation
for the phase estimation and the quantum counting,
called the {\it semi-classical}
implementation~\cite{PP00efficient_factorization}.


The rest of this paper is organized as follows:
In Section~\ref{sec:preliminaries},
we begin by explaining the quantum counting and
the depolarizing channel used in this paper,
and implementing two quantum counting circuits.
In Section~\ref{sec:analysis_counting},
we analyze the decoherence on the quantum counting.
We first consider analysis model for the decoherence,
and then investigate probability distributions on the quantum counting
in the presence of the decoherence
on the first register and the second register.
In Section~\ref{sec:phase_estimation},
we also discuss influences of decoherence on the phase estimation
and on the semi-classical implementation.
In Section~\ref{sec:conclusion}, we summarize our results.
Finally,
we give some detailed calculations of probability distributions
in the case that the decoherence error occurs on the first register
and the second one in~\ref{sec:appendix_first}
and \ref{sec:appendix_second}, respectively.


\section{Definitions, notations, and decoherence model}
\label{sec:preliminaries}

First of all,
we describe definitions and notations of the quantum counting
used in this paper,
and explain our decoherence model.
We also implement two quantum counting circuits
for revealing robustness against decoherence.

\subsection{Quantum counting}
\label{sec:counting}

Suppose that there are unordered $N:=2^n$ elements
and a function $f:\{0, \dots, N-1\} \rightarrow \{0,1\}$
is given as an {\itshape oracle} (a black box).
Let $\cI$ be a set of elements which satisfy $f(x)=1$
of $N$ elements,
that is,
\[
 f(x) = \cases{
  1 & ($x \in \cI$) \\
  0 & ($x \notin \cI$)
 },
\]
and $t := |\cI|$.
The goal of a quantum counting algorithm
is to estimate the number $t$
of solutions of the oracle $f$~\cite{BHT98counting}.

In order to estimate $t$,
the quantum counting estimates the phase of
a unitary operator $G$, called {\it Grover operator\/},
by using the quantum Fourier transform.
Let $\ket{x} \in \cH := (\bC^2)^{\tensor n}$,
where $(0 \le x \le 2^n-1)$ be
quantum states which represent $2^n$ elements.
Grover operator $G$ is defined as 
$G:=U_2U_1$ on $\cH$,
where
$U_1 := \sum_x(-1)^{f(x)}\mixed{x}{x},
U_2 := 2\mixed{s}{s}-\mathbf{1}_{N},
\ket{s} := \amp{N}\sum_{i=0}^{N-1}\ket{i}$.
The operator $G$ can be rewritten as a rotation
on two-dimensional space.
We divide the Hilbert space $\cH$
into the ``good'' space
$\cH_g := \varstackrel{x \in \mathcal{I}}{\mathrm{span}} \{ \ket{x} \}$
and the ``bad'' space
$\cH_b := \varstackrel{x \not\in \mathcal{I}}{\mathrm{span}} \{ \ket{x}
\}$
and define two orthonormal states on each space:
\begin{equation}
 \eqalign{
 \ket{b} := &\frac{1}{\sqrt{N-t}} \sum_{x \not\in {\mathcal I}} \ket{x}
 \in {\mathcal H}_b, \nonumber \\
 \ket{g} := &\frac{1}{\sqrt{t}} \sum_{x \in {\mathcal I}} \ket{x}
 \in {\mathcal H}_g.}
\end{equation}
The two-dimensional vector space spanned by
the bases $\ket{b}$ and $\ket{g}$ is called
{\it Grover space\/}.
The Grover operator $G$ can be represented
on the Grover space as follows:
\begin{equation}
 G \equiv
  \left(
  \begin{array}{cc}
   \cos \theta & -\sin \theta \\
   \sin \theta & \cos \theta
  \end{array}
  \right),
\end{equation}
where $\sin (\theta / 2) := \sqrt{t/N}$.

\begin{figure}[t]
 \begin{center}
  \scalebox{1.00}{\includegraphics{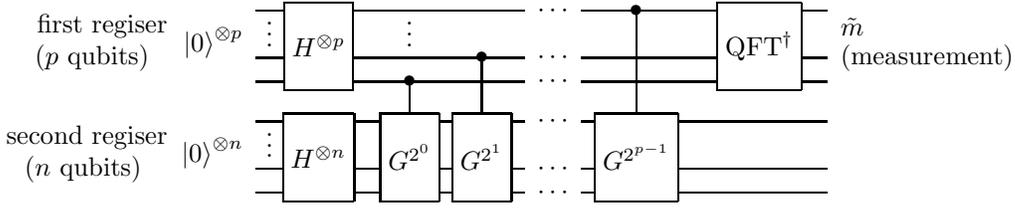}}
 \end{center}
 \caption{\label{fig:counting_circuit}Circuit for the quantum counting}
\end{figure}

Figure~\ref{fig:counting_circuit} shows
a circuit for the quantum counting.
We refer to the upper $p$ qubits in this figure as a \textit{first register} and
the lower $n$ qubits as a \textit{second register}.
The quantum counting algorithm is composed of the following five stages.
\begin{enumerate}
 \item Prepare an initial state $\ket{0}^{\tensor (p+n)}$.
       \label{enu:counting_1}
 \item Apply the Hadamard transformation $H$ to all qubits:
       $\amp{P} \sum_{m=0}^{P-1} \ket{m} \tensor \ket{s},$
       where $P := 2^p$, $\ket{m}$ and $\ket{s}$ belong to
       the first and second register, respectively.
       \label{enu:counting_2}
 \item Apply controlled-$G$ to the second register
       according to the first register $\ket{m}$.
       \[
       \amp{P} \sum_{m=0}^{P-1} \ket{m} \tensor G^m \ket{s}.
       \]
       \label{enu:counting_3}
 \item Apply the inverse Fourier transform to the first register.
       \begin{eqnarray}
	 \frac{1}{P} \sum_{m'=0}^{P-1}
	 \ket{m'} \mbox{} \otimes \sum_{m=0}^{P-1}
	 \exp \left(2 \pi \rmi \frac{mm'}{P} \right) G^m \ket{s} \nonumber \\
	\lo=
	 \frac{1}{\sqrt{2}} \sum_{m'=0}^{P-1} e^{\frac{P-1}{P} \pi{}\rmi{}m'}
	 \ket{m'}
	 \nonumber \\
	\otimes
	 \left[
	  e^{\pi{}\rmi{}f} \sinNoP{(m'+f)} \ket{+}
	  + e^{-\pi{}\rmi{}f} \sinNoP{(m'-f)} \ket{-}
	 \right], \label{eq:counting_noerror}
       \end{eqnarray}
       where $f:=P\theta / 2\pi$,
       $\ket{\pm} := \amp{2}(\ket{b} \mp \rmi\ket{g})$.
       \label{enu:counting_4}
 \item Measure the first register and obtain
       $\tilde{m}$ for a good estimator of $f$. 
       \label{enu:counting_5}
\end{enumerate}


In order to determine the number of solutions
from the measurement result $\tilde{m}$,
we calculate $\tilde{t} := N \sin^2 (\tilde{\theta}/2) = N \sin^2 (\pi
\tilde{m} / P )$.
Since the probability distribution of
Equation~(\ref{eq:counting_noerror}) 
has peaks at
$\tilde{m} \simeq f, P-f (P \gg 1)$,
we obtain the output $\tilde{t}$ with high probability,
which is an approximation of $t$ for the quantum counting.

In this algorithm, we have to fix the parameter $P$
that determines the precision.
At first, by running this algorithm with setting $P$ at $\sqrt{N}$,
we obtain an approximated $\tilde{t}$ such that
$|t-\tilde{t}| < 2\pi\sqrt{t} + \pi^2$.
Then, by running it again with setting $P$ at $20\sqrt{\tilde{t}N}$,
we obtain \textit{new} estimation $\tilde{t}$ as a more precise result.
It is guaranteed to obtain $t$ with probability
at least $8/\pi^2$~\cite{BHT98counting}.
It follows that the time needed for the quantum counting
is $\Or(2^p)=\Or(\sqrt{tN})$.

\begin{figure}
 \begin{minipage}{0.49\textwidth}
  \begin{center}
   \includegraphics[width=\textwidth]{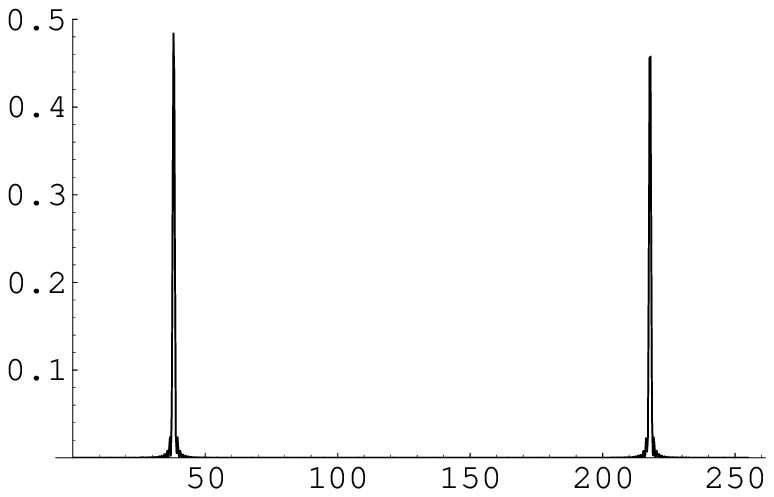}
   
   \vspace{-1.0mm}
   $m'$
   \vspace{-1mm}
  \end{center}
  \caption{\label{fig:no_error_m}%
  The probability distribution of
  the measurement result $m'$.
  Peaks appear near $f \simeq 38$ and $2^p-f\simeq 218$.
  }
 \end{minipage}
 \begin{minipage}{0.49\textwidth}
  \begin{center}
   \includegraphics[width=\textwidth]{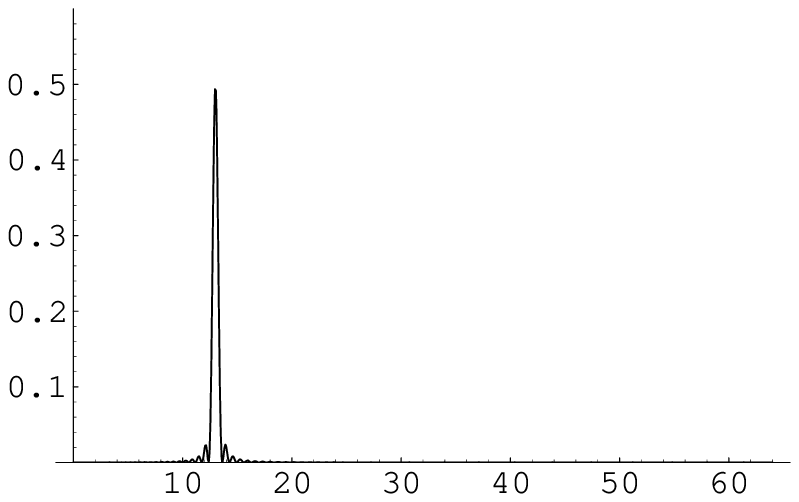}
   
   \vspace{-1.0mm}
   $t'$
   \vspace{-1mm}
  \end{center}
  \caption{\label{fig:no_error_t}The probability distribution of
  the output $t'$. \\ \mbox{} \\}
 \end{minipage}
\end{figure}

In Figure~\ref{fig:no_error_m},
we show an example of the probability distribution of $\ket{m'}$ in
Equation~(\ref{eq:counting_noerror})
for the quantum counting,
by setting $p=6, \ n=8, \ \cI=\{0,\dots,12\}$. 
We also show a probability distribution
of the corresponding output $t'$
in Figure~\ref{fig:no_error_t}.
The probability distribution
of $m'$ has two peaks, and
that of $t'$ has single peak
at $t$ that is the desired number of solutions.

\subsection{Decoherence model}
\label{sec:errormodel}

In order to analyze the decoherence,
we need to model decoherence error.
One of useful error models
on classical information
is the binary symmetric channel,
which flips a bit
with probability $p$
and leaves it alone with probability $1-p$.
We consider the following quantum decoherence model
analogous to the binary symmetric channel,
which is often used
for analyzing error correcting codes.

If we do not know anything about properties of errors
that the quantum system suffers from,
it is one of reasonable error models
that quantum states are disturbed into maximally mixed state
as time goes on.
In this paper, we assume that errors occur as local \textit{depolarizing channel}.
\begin{definition}[depolarizing channel~\cite{NC00computation}]
 \label{def:depolarizing}
\begin{eqnarray}
  \rho &\rightarrow (1-d)\rho + d\cdot\frac{I}{2} \nonumber \\
 &= (1-d)\rho  + d \cdot
 \frac{\sigma_0\rho\sigma_0 + \sx\rho\sx + \sy\rho\sy + \sz\rho\sz}{4}
 \label{eq:depolarizing},
\end{eqnarray}
where $\rho$ is a density matrix on $\mathbb{C}^2$, 
 $\sigma_0$ is the identity operator,
 and $\sigma_x, \sigma_y, \sigma_z$ are {\itshape Pauli} matrices.
\end{definition}

We apply this time-discretized error to each qubit at each unit time
regardless of the existence of a quantum gate on the qubit.
Since Equation~(\ref{eq:depolarizing}) is represented
by summation of classical events,
we simulate this channel by applying
$\sx, \sy, \sz$, and $\sigma_0$ to each qubit
with probability $d/4$;
otherwise the state is left untouched.
On our numerical calculations, we take an average of experiments
by using only pure states.

\subsection{Two implementations for the quantum counting}
\label{sec:two_implementations}

\begin{figure}[t]
 \begin{minipage}{0.49\textwidth}
  \begin{center}
   \scalebox{1.2}{
   \includegraphics{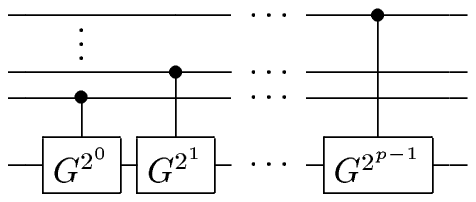}
   }
   
   \vspace{1mm}
   \mbox{(a)}
  \end{center}
 \end{minipage}
 \begin{minipage}{0.49\textwidth}
  \begin{center}
   \scalebox{1.2}{
   \includegraphics{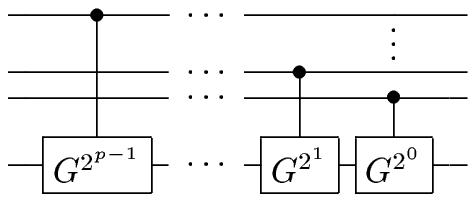}
   }
   
   \vspace{1mm}
   \mbox{(b)}
  \end{center}
 \end{minipage}
 \caption[Two implementations for the quantum counting]{
 Two implementations of controlled-$G^m$ operations.
 The ordering~(a) and (b) are called
 the {\itshape ascending-order} and the {\itshape descending-order}, respectively.
 Two circuits are {\itshape equivalent} if no error occurs.
 }
 \label{fig:two_implementations}
\end{figure}

Generally speaking,
unitary operations in circuit are not commutative.
In the quantum counting circuit, however,
all controlled-$G$ gates are commutative.
Figure~\ref{fig:two_implementations}
shows examples of {\it equivalent} implementations
for the quantum counting, i.e.{}\ 
probability distributions on two implementations
are completely the same in no decoherence case.
One of difficulties in the analysis of decoherence
is that we can not exchange the operators that are commutative
in the system which is affected by decoherence.

One of our aims in this paper is to reveal
how different influences of the decoherence
on {\itshape equivalent} circuits are
and
which implementation is more robust against the decoherence.
In order to investigate it,
we implement two typical circuits for the quantum counting.
\begin{definition}[two implementations for the quantum counting]
 We define two quantum counting circuits
 with the following ordering.
 \begin{description}
  \item[ascending-order]
	     from the controlled-$G^{2^0}$ operation
	     in Figure~\ref{fig:two_implementations}~(a).
  \item[descending-order]
	     from the controlled-$G^{2^{p-1}}$ operations
	     in Figure~\ref{fig:two_implementations}~(b).
 \end{description}
\end{definition}
%
%
The descending-order implementation is
especially needed for
an efficient implementation of the quantum counting
and the phase estimation algorithm,
which reduces the number of qubits on the first register
to one.
We discuss these efficient implementations
in Section~\ref{sec:analysis_semi-classical} later.


\section{Decoherence on the quantum counting}
\label{sec:analysis_counting}

In this section, we analyze influences
of the decoherence on the quantum counting.
We begin by explaining our analysis model
and then investigate the decoherence
on the first and the second register
in two quantum counting circuits
with the ascending-order and the descending-order.
Finally, we discuss
the robust implementation for the quantum counting
against the decoherence.


\subsection{Analysis model for the decoherence}
\label{sec:analysis_method}

The probability that
the decoherence error occurs is considered to
increase in proportion to the product
of execution time and the number of qubits.
As stated in Section~\ref{sec:counting},
the quantum counting is performed in five stages:
(\ref{enu:counting_1}) preparing an initial state,
(\ref{enu:counting_2}) applying the Hadamard transformation,
(\ref{enu:counting_3}) applying controlled-$G$ operations,
(\ref{enu:counting_4}) applying the inverse quantum Fourier transform,
and (\ref{enu:counting_5}) measurements.
The time needed for each stage is as follows:
The stage~(\ref{enu:counting_1}) and (\ref{enu:counting_5}) are considered
to be one step.
Application of the Hadamard transformations
on the stage (\ref{enu:counting_2})
takes $O(1)$ and
application of the inverse quantum Fourier transform
on the stage (\ref{enu:counting_4})
takes at most $O(p^2)$.
In contrast, application of controlled-$G$ operations
on the stage~(\ref{enu:counting_3}) takes $2^p-1$, which is exponential to $p$.
It follows that the decoherence error happens on the stage~(\ref{enu:counting_3})
with exponentially higher probability
than on other stages.
On our analyses, we restrict
the position of the error only on the stage~(\ref{enu:counting_3}).

Suppose that the error rate $d$ is small enough $(d \ll 1)$,
then influences of the decoherence are approximated
by the first order term of $d$.
This means that the depolarizing channel is applied only once.
%
%
Since the quantum counting has two registers,
the register where the depolarizing channel is applied
is either of the first register or the second one.
%
%

Under these conditions,
we calculate probability distributions on the quantum counting,
and give some properties of the decoherence on two registers.
Moreover, we compare influences of the decoherence
on the ascending-order and the descending-order implementations.


\subsection{Decoherence on the first register}
\label{sec:analysis_first}

\begin{figure}[t]
 \begin{center}
  \includegraphics{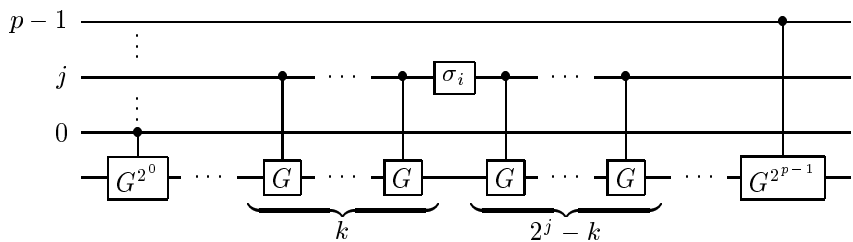} \\
 \end{center}
 \caption[Position of $\sigma_i$ error on the first register]{
 $\sigma_i$ error occurs
 on the $j$-th qubit of the first register
 after application of controlled-$G^k$ operations with the $j$-th control qubit.
 }
 \label{fig:position_error1st}
\end{figure}

We analyze the decoherence on the first register
by considering the case
that the error based on the depolarizing channel
occurs once on the first register.
Through the depolarizing channel,
$\sigma_x, \sigma_y, \sigma_z$ errors
and identity operator $\sigma_0$ occur
with the same rate
from Equation~(\ref{eq:depolarizing}).
We first deal with the ascending-order circuit
in Figure~\ref{fig:two_implementations}~(a).

We investigate influences of the decoherence
by calculating probability distributions on the quantum counting.
For calculations,
we need to represent the position of $\sigma_i$ error.
As stated above,
we already restrict the position of the decoherence
on the state~(\ref{enu:counting_3}).
Let $j$ and $k$ be integers such that
the error occurs
\begin{itemize}
 \item on the $j$-th qubit of the first register,
 \item after application of controlled-$G^k$ operations
       with the $j$-th control qubit,
\end{itemize}
where $0 \le j \le p-1, \ 0 \le k \le 2^j$.
These parameters are sufficient for determining
where and when $\sigma_i$ error occurs.
The case that
the error occurs on the $j$-th qubit
before(after) controlled-$G^{2^j}$
can be represented by $k=0 (k=2^j)$ respectively
since our decoherence model is local.
Figure~\ref{fig:position_error1st} shows
the position of $\sigma_i$ error on our analyses.
The total number of applications
of controlled-$G$ operations
before the error
is $2^j+k-1$.

We first focus on the position of peaks
in a probability distribution on the quantum counting
under the decoherence.
If no error occurs,
the probability distribution has
only two correct peaks
near $\tilde{m} \simeq f, 2^p-f$,
as shown in Equation~(\ref{eq:counting_noerror}).
 Let $Prob^{(i,j,k)} (m')$ be
 the probability to obtain $m'$ as a measurement result
 in the case of the above position of $\sigma_i$ error.
 By calculation $Prob^{(i,j,k)} (m')$
 in~\ref{sec:appendix_first},
 we have
 \begin{eqnarray}
  \sum_{i=0, x, y, z} Prob^{(i, j, k)}(m') & = & \hspace{31mm} \nonumber \\
  & & \hspace{-4cm}
   \Biggl[
   \frac{\sin \left\{ \pi ( m' + f ) \right\} }
   {2^{p-j-1} \sin \left\{ \pi/ 2^{p-j-1}( m' + f ) \right\} } \nonumber
   \mbox{} \times
   \frac{2^{p-j} \sin \left\{ \pi / 2^{p-j} ( m' + f ) \right\} }
   {2^{p}   \sin \left\{ \pi / 2^{p} ( m' + f )   \right\} }
   \Biggr]^2 \nonumber \\
  & & \hspace{-4cm}
   \mbox{}+
   \Biggl[
   \frac{\sin \pi ( m' - f )}
   {2^{p-j-1} \sin \left\{ \pi/ 2^{p-j-1}( m' - f ) \right\} }
   \mbox{} \times
   \frac{2^{p-j} \sin \left\{ \pi / 2^{p-j} ( m' - f ) \right\} }
   {2^{p}   \sin \left\{ \pi / 2^{p} ( m' - f )   \right\} }
   \Biggr]^2. \label{eq:prob_ijk}
 \end{eqnarray}
 This equation has two strong peaks at $m'\simeq f, -f\equiv 2^p-f$
 that are the same as the correct peaks
 and weak peaks at a distance of $\pm 2^{p-j-1}$ from the strong peaks,
 which is caused by errors.
\begin{proposition}
 \label{thm:positions_wrong_first}
 The probability distribution
 related to the first order term of
 the error rate
 on the quantum counting
 mainly has wrong peaks
 at a distance of the power of two from $\tilde{m} \simeq f, 2^p - f$
 if the depolarizing channel is applied
 on the first register.
\end{proposition}

\begin{figure}[t]
 \begin{center}
  \begin{minipage}{0.45\linewidth}
   \begin{center}
    \hspace{-10mm}
    \scalebox{0.87}{\includegraphics{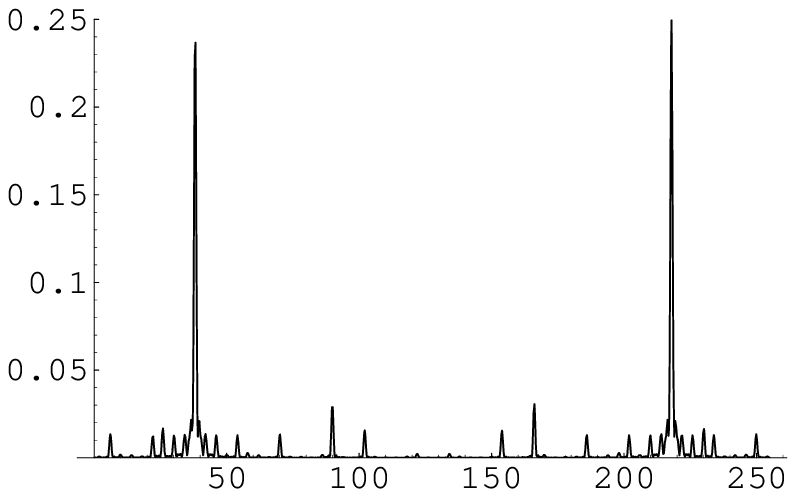}} \\
    \vspace{-2mm}\hspace{4mm}\scalebox{1.00}{$m'$}\\
    \mbox{\hspace{3mm} (a)}
   \end{center}
  \end{minipage}
  \begin{minipage}{0.45\linewidth}
   \begin{center}
    \scalebox{1.0}{\includegraphics[angle=270,width=\textwidth]{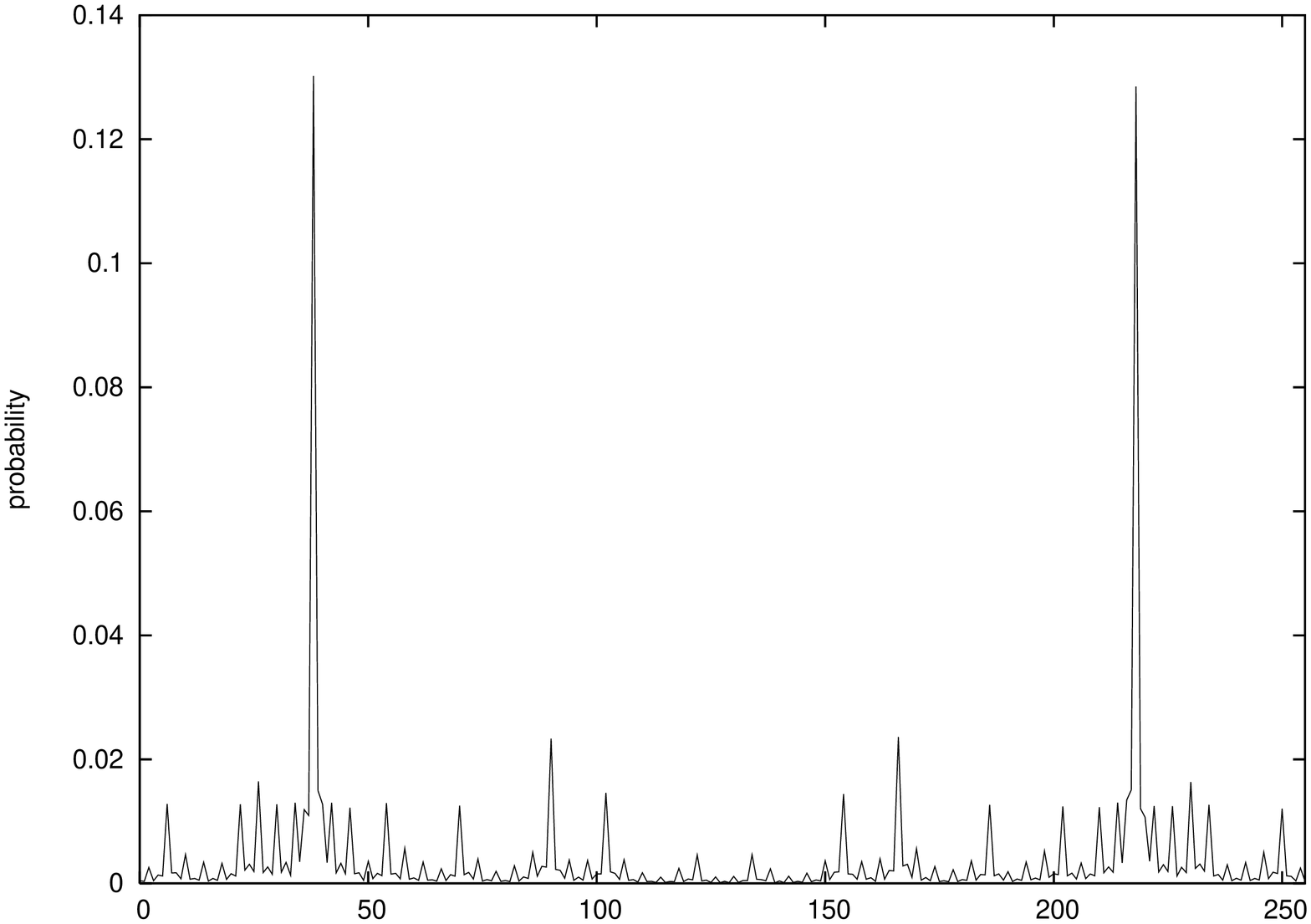}}

    \hspace{1mm}
    \vspace{0.5mm}
    measurement result $\tilde{m}$ \\
    \hspace{1mm}(b)
   \end{center}
  \end{minipage}
  \caption[Graphs of measurement result in the first register case]{
  The probability distributions of measurement results
  with the same parameters in Figure~\ref{fig:no_error_m}.
  (a):~$\frac{1}{4p} \sum_{j=0}^{p-1} \sum_{i=0,x,y,z}
  Prob^{(i,j,k)}(m')$.
  (b): the average of numerical calculations $10^5$ trials
  where the error rate $d=4 \times 10^{-3}$.
  }
  \label{fig:err1}
 \end{center}
\end{figure}

%
%

We show the graph of
$\frac{1}{4p} \sum_{i=0,x,y,z} \sum_{j=0}^{p-1} Prob^{(i,j,k)}(m')$
in Figure~\ref{fig:err1}~(a),
which means the average of probability distributions
in all error cases.
We also show the graph of numerical calculations
in Figure~\ref{fig:err1}~(b).
We did the experiments on Quantum Computation Simulation System
(QCSS)~\cite{NMI02simulator}
and took the average of $10^5$ trials.
On the numerical calculations,
we set the error rate $d= 4\times 10^{-3}$
so that the decoherence errors
based on the depolarizing channel
occur on the first register about twice on each trial.

Proposition~\ref{thm:positions_wrong_first}
follows the number of wrong peaks caused by the decoherence.

\begin{claim}
 The number of main wrong peaks is $O(p)$
 in probability distribution
 related to the first order term of
 the error rate
 if the depolarizing channel is applied 
 on the first register.
\end{claim}

%

The probability distribution
$\sum_{i=0,x,y,z} Prob^{(i,j,k)}(m')$ does not
depend on $k$ that determines {\it depth} of $\sigma_i$ error,
i.e.{}\ the time when the error occurs.

\begin{proposition}
 \label{thm:independence_first}
 The probability distribution
 related to the first order term of
 the error rate
 on the quantum counting
 is independent of depth of error
 if the depolarizing channel is applied
 on the first register.
\end{proposition}

Proposition~\ref{thm:independence_first}
means that there is no difference
of influences of the decoherence
between on the ascending-order circuit
and on the descending-order one for the quantum counting,
since controlled-Grover operators are
commutative each other.

\begin{proposition}
 \label{thm:independence_implementation}
 Probability distributions
 related to the first order term of
 the error rate
 on the quantum counting
 are independent of the ordering of
 application of controlled-$G$ operations
 if the depolarizing channel is applied
 on the first register.
\end{proposition}


\subsection{Decoherence on the second register}
\label{sec:analysis_second}


We then deal with the case
that the decoherence error occurs once on the second register
in the quantum counting.
%
In Section~\ref{sec:analysis_first},
we treat the error on the first register,
which consists of control gates.
In that case, error affects only the number of application $m$
of controlled-$G^m$.
On the other hand, the error on the second register
modifies the state on which Grover operator acts.
Because the states $\ket{b}$ and $\ket{g}$ depend
on the quantum oracle,
we can not specify how the second register is disturbed
by the decoherence.
We need to consider the disturbance
and action of Grover operator on a disturbed second register.

We first show action of Grover operator
on an arbitrary quantum state
in order to deal with the operator on a disturbed
second register.
Any quantum state $\ket{\phi} \in \mathcal{H}$ is decomposed
as follows
by means of Gram-Schmidt orthogonalization:
\begin{equation}
 \ket{\phi} :=
  \fourGroverstate{\phi},
  \label{eq:phi}
\end{equation}
where $\up, \vp, \uep, \vep \in \mathbb{C}$,
$\ketebp \in \mathcal{H}_b, \ \ketegp \in \mathcal{H}_g$, and
$\ketebp$ and $\ketegp$ are determined
such that $\inner{b}{\ebp} = \inner{g}{\egp} = 0$.
By definition of Grover operator,
we obtain the following lemma.
\begin{lemma}
 \label{lem:G_four_bases}
 For any quantum state $\ket{\phi}$,
 Grover operator $G$ can be rewritten as:
 \begin{equation}
  G \equiv
   \left(
    \begin{array}{cccc}
    \cos \theta & -\sin \theta & 0 & 0 \\
    \sin \theta & \cos \theta & 0 & 0 \\
    0 & 0 & -1 & 0 \\
    0 & 0 & 0 & 1
    \end{array}
   \right),
   \label{eq:G_four_bases}
 \end{equation}
 on four-dimensional space
 spanned by the basis states
 $\ket{b}, \ket{g}, \ketebp$, and $\ketegp$,
 which satisfy
  $\ketebp \in \mathcal{H}_b, \ \ketegp \in \mathcal{H}_g, 
  \inner{b}{\ebp} = \inner{g}{\egp} = 0.$
\end{lemma}

Before disturbance, the second register is a superposition
of the states $\ket{b}$ and $\ket{g}$.
If some error occurs on the register,
two states are disturbed into
\begin{eqnarray}
 \ket{b} &\rightarrow 
  \fourGroverstate{b}, \nonumber \\
 \ket{g} &\rightarrow 
  \fourGroverstate{g},
  \label{eq:disturbed_bg}
\end{eqnarray}
satisfying
 \begin{eqnarray}
  \ketebb, \ketebg \in \mathcal{H}_b, \
  \ketegb, \ketegg \in \mathcal{H}_g, \nonumber \\
  \inner{b}{\ebb} = \inner{b}{\ebg} =
  \inner{g}{\egb} = \inner{g}{\egg} = 0.
  \label{eq:six_properties}
 \end{eqnarray}
%
\begin{lemma}
 \label{lem:Grover_six_bases}
 Any disturbed second register by the decoherence
 can be represented by the superpositions of
 $\ket{b}, \ket{g}, \ketebb, \ketebg, \ketegb$, and $\ketegg$,
 satisfying Equation~(\ref{eq:six_properties}).
 Action of Grover operator is a rotation by $\theta$
 on two-dimensional space spanned by $\ket{b}$ and $\ket{g}$,
 by $\pi$ on $\ketebb$ and $\ketebg$,
 and by $0$ on $\ketegb$ and $\ketegg$.
\end{lemma}

%


\begin{figure}[t]
 \begin{center}
  \scalebox{1.0}{ \includegraphics{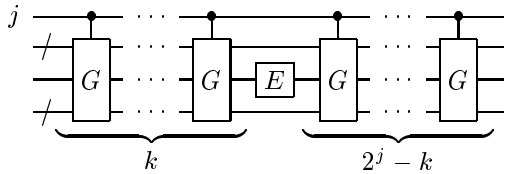} }
 \end{center}
 \caption[Position of some error on the second register]{Some error $E$ occurs on the second register
 after application of controlled-$G^k$ operations with the $j$-th
 control qubit
 $(0 \le j \le p-1, 0 \le k \le 2^j)$. }
 \label{fig:circuit_error2nd}
\end{figure}

Like the case of the first register,
we consider that
some error $E$, not necessarily the depolarizing channel,
is applied on the second register
after application of controlled-$G^k$ operations
with the $j$-th control qubit,
as shown in Figure~\ref{fig:circuit_error2nd}.
\begin{figure}[Ht]
 \begin{center}
  \begin{minipage}{0.45\linewidth}
   \begin{center}
    \hspace{-2mm}
    \scalebox{1.0}{\includegraphics[angle=270,width=\textwidth]{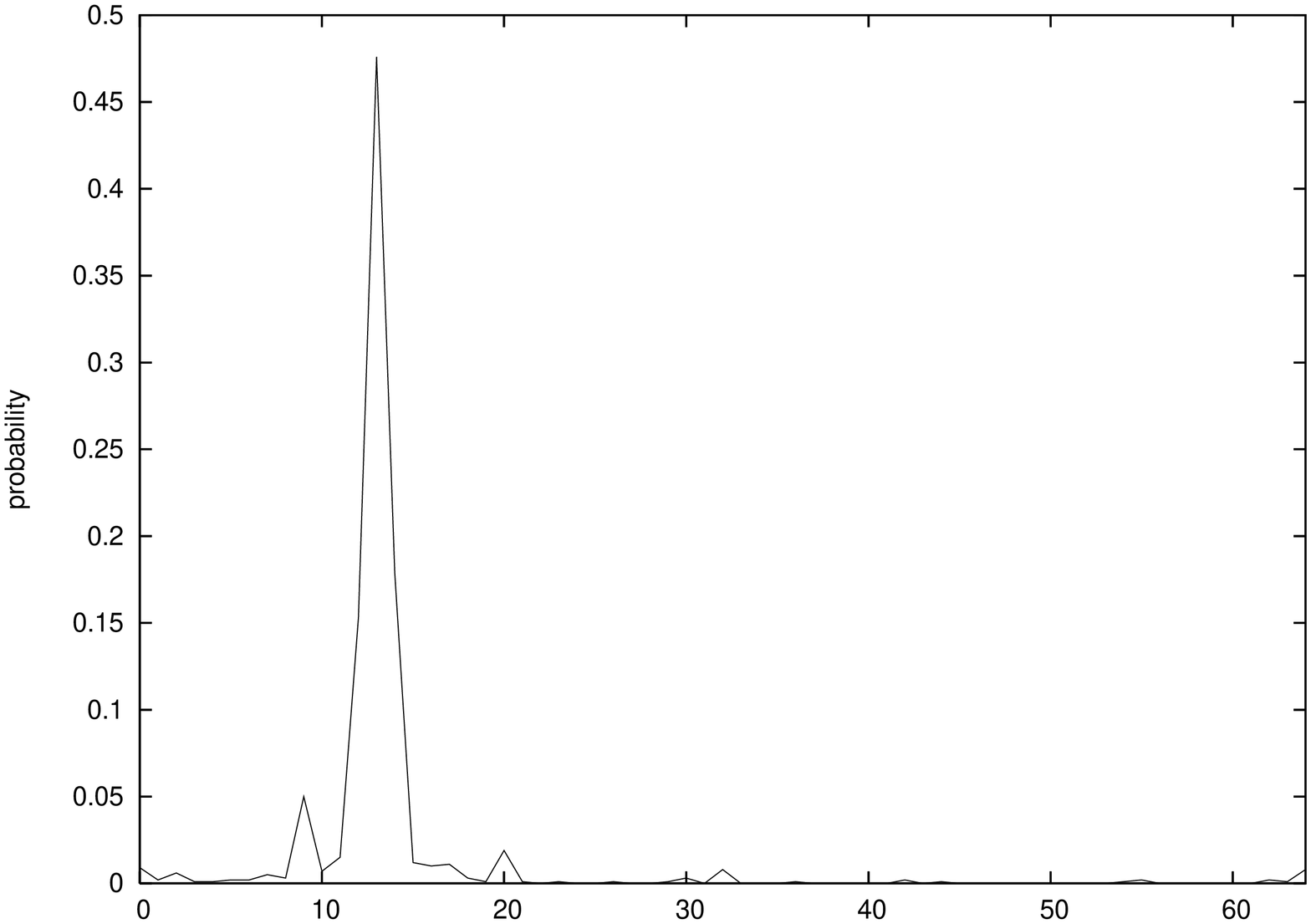}}

    output $\tilde{t}$ \\
    (a)
   \end{center}
  \end{minipage}
  \begin{minipage}{0.45\linewidth}
   \begin{center}
    \scalebox{1.0}{\includegraphics[angle=270,width=\textwidth]{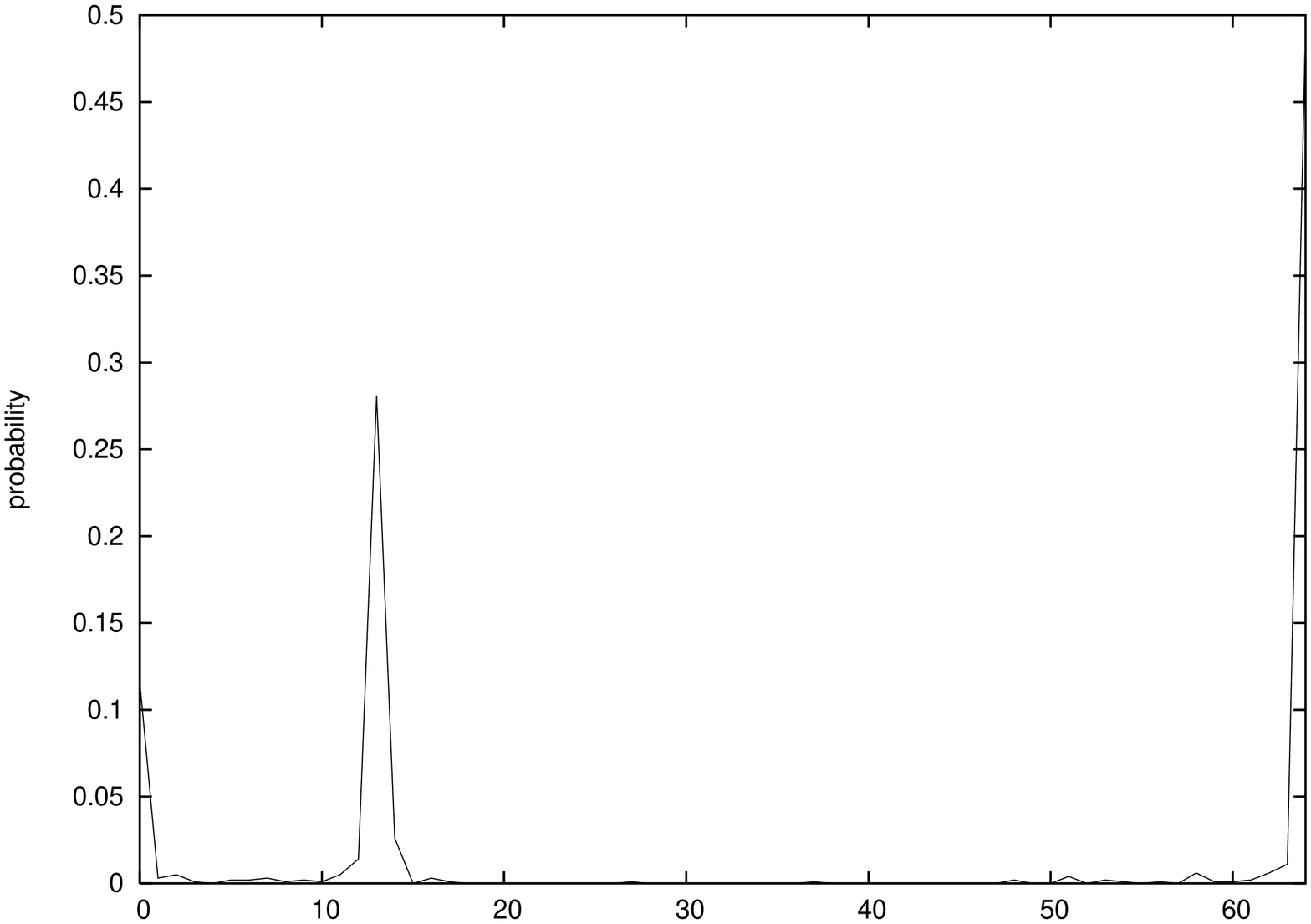}}

    \hspace{1.5mm}
    output $\tilde{t}$ \\
    \hspace{1.5mm}(b)
   \end{center}
  \end{minipage}
  \vspace{1mm}
  \caption[The output in the second register case]{
  The probability distributions of the output $\tilde{t}$
  by numerical calculations
  $10^3$ trials with the error rate $d=4\times 10^{-3}$.
  (a): on the ascending-order circuit,
  (b): on the descending-order circuit.
  }
  \label{fig:err2_outputs}
 \end{center}
\end{figure}
%
%
%
In this case, 
a probability distribution
before measurement
on the ascending-order circuit
has peaks near
$m' \simeq f, 2^p-f$ that are
the same as ones in no error case,
as detailed in \ref{sec:appendix_second}.
On the other hand,
a probability distribution
on the descending-order circuit
has peaks not only near $m' \simeq f, 2^p-f$
but also at $m' = 0, 2^p/2$
with high probability,
independently of the quantum oracle.
By calculating an output $t'=N\sin^2 (\pi m'/2^p)$ of the quantum counting,
we obtain the following proposition.
\begin{proposition}
 \label{thm:position_outputs_second}
 The following wrong outputs of the quantum counting
 related to the first order term of
 the error rate
 are obtained with high probability
 if some decoherence error occurs on the second register:
 \begin{itemize}
  \item Wrong outputs near $t$
	in the ascending-order case.
  \item Wrong outputs 0 and $N$
	in the descending-order case,
	independently of the quantum oracle.
 \end{itemize}
\end{proposition}

We show two graphs of outputs of
the quantum counting
with the ascending-order and the descending-order
by numerical calculations
in Figure~\ref{fig:err2_outputs}~(a) and (b),
respectively.
These experiments were done $10^3$ trials
with the same conditions as the first register case
except that the decoherence error occurs on the second register.

%
%
%
 The difference of positions of wrong peaks
 between two quantum counting circuits
 can be intuitively considered as follows:
 If a decoherence error occurs
 on the second register,
 influences of the error propagate to
 the first register by controlled-$G$ operators.
 On the ascending-order circuit,
 $G$s are applied from the controlled-$G^{2^0}$
 with the $0$th control qubit
 corresponding to the MSB of a measurement result.
 The influences therefore propagate
 the $0$th control qubit(MSB) to $(p-1)$th control qubit(LSB).
 Since application of controlled-$G^m$ operations
 with the LSB needs
 more time exponentially than with the MSB,
 decoherence error occurs
 with exponential higher probability
 on controlled-$G^m$ operations with low control qubits.
 It follows that influences of the error
 propagate to only low control qubits.
 On the other hand, in the descending-order case,
 the influences propagate from
 the LSB to the MSB of the first register
 because of reversed ordering of application of
 controlled-$G$ operations.
 Therefore not only low control qubits but also
 high qubits are affected by the decoherence.

We finally consider the probability to obtain
the correct output for the quantum counting.
Peaks in the probability distribution on the descending-order circuit
are distributed to four peaks
whereas the probability distribution on the ascending-order circuit
has only the correct two peaks.

\begin{proposition}
 \label{thm:higher_probability_second}
 The correct output of the quantum counting
 is obtained
 with higher probability
 on the ascending-order circuit
 than on the descending-order circuit
 if the decoherence error occurs on the second register.
\end{proposition}



\subsection{Robust implementation against the decoherence}
\label{sec:analysis_robust}

Proposition~\ref{thm:independence_implementation}
shows that the probability to obtain the correct output
on the ascending-order is
the same as the probability on the descending-order one
in the case of decoherence on the first register.
Proposition~\ref{thm:higher_probability_second}
together with the proposition
states
the following robustness against the decoherence on two registers
in the quantum counting.


\begin{claim}
 The ascending-order implementation
 for the quantum counting
 is more robust against the decoherence
 than the descending-order implementation.
\end{claim}

%


\section{Discussion on phase estimation algorithms}
\label{sec:phase_estimation}

In this section,
we extend our results with respect to
robust implementation against the decoherence
to the phase estimation.
We also discuss
robustness of
semi-classical implementation
against the decoherence.

\subsection{Decoherence on phase estimation algorithms}
\label{sec:decoherence_phase_estimation}

\begin{figure}[t]
 \begin{center}
  \scalebox{0.90}{\includegraphics{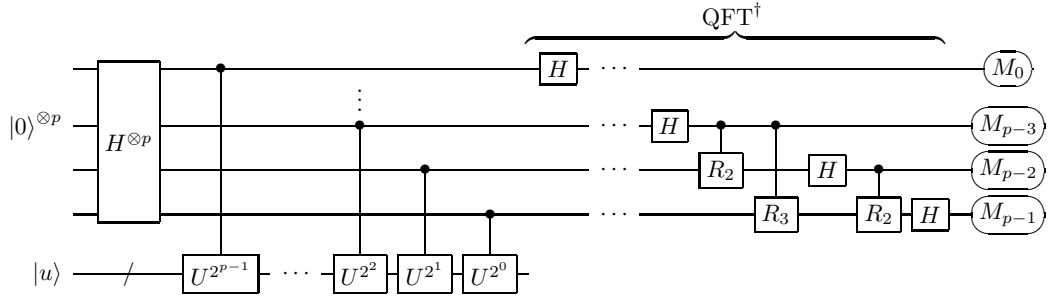}}
  \end{center}
 \caption[Circuit for the phase estimation]{Circuit for the phase estimation
 with the descending-order.
 Controlled rotations $R_j$ are defined as
 $R_j :=${\scriptsize
	 $\left( \begin{array}{cc} 1 & 0 \\ 0 & \phi_j \end{array} \right)$}
 with $\phi_j := e^{-2\pi{}i/2^j}$.}
 \label{fig:circuit_phase_estimation}
\end{figure}

Phase estimation is one of key quantum algorithms,
used in Shor's factoring~\cite{Shor94factorization}.
Suppose that a unitary operator $U$ has
an eigenvector $\ket{u}$
with eigenvalue $e^{2\pi{}i\varphi}$,
where the value of $\varphi$ is unknown.
The goal of the phase estimation is to
find the phase $\varphi$.
A circuit for the phase estimation is shown
in Figure~\ref{fig:circuit_phase_estimation}.
The phase estimation is performed
by application of controlled-$U$ operations
to the second register prepared
to the corresponding eigenvector $\ket{u}$ initially.
Like the quantum counting,
the phase estimation has such {\itshape equivalent} implementations
as the ascending-order circuit and the descending-order circuit,
by changing ordering of application of controlled-$U$ operations
instead of controlled-$G$ operations in the quantum counting.
\begin{claim}
 The phase estimation can
 find the desired phase
 on the ascending-order circuit
 with higher probability
 than on the descending-order one
 in the presence of decoherence on the second register.
\end{claim}

As mentioned in Subsection~\ref{sec:analysis_second},
influences of the decoherence error on the second register
in the quantum counting
propagate only to lower qubits of the first register
with exponentially high probability on the ascending-order circuit,
whereas these influences propagate to
higher qubits of the first register on the descending-order circuit.
This propagation of decoherence
is applicable not only to the quantum counting
but also to the phase estimation,
though positions of wrong peaks caused by the decoherence
are determined by a unitary operator $U$ on the phase estimation.

\subsection{Decoherence on efficient implementations for the phase estimation}
\label{sec:analysis_semi-classical}

\begin{figure}
 \begin{center}
  \scalebox{0.9}{
  \includegraphics{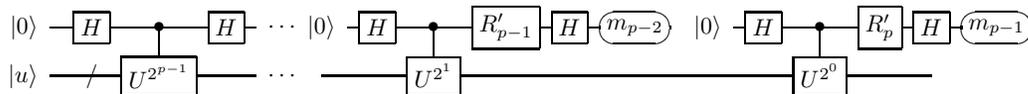}
  }
 \end{center}
 \caption[Semi-classical circuit for the phase estimation]
 {Semi-classical circuit for the phase estimation.
 $R'_j$ is defined with the results of previous measurements:
 $R'_j =${\scriptsize $\left(\begin{array}{cc} 1&0\\ 0&\phi'_j
			     \end{array} \right)$}
 with $\phi'_j := e^{-2\pi{}i\sum_{k=2}^{j} m_{j-k}/2^k }$.}
 \label{fig:circuit_semi_classical}
\end{figure}

An efficient implementation for the phase estimation,
called the {\itshape semi-classical}
implementation, was demonstrated
by Parker et al.~\cite{PP00efficient_factorization}.
We show the semi-classical circuit for the phase estimation
in Figure~\ref{fig:circuit_semi_classical},
corresponding to the circuit
in Figure~\ref{fig:circuit_phase_estimation}.
On this circuit, inverse quantum Fourier transform
and measurements are applied
as fast as possible
so that only single qubit is required for the first register.
By using semi-classical technique,
Beauregard~\cite{Beauregard022nplus3}
reduced the number of qubits
for the factoring about half as implementations
in Figure~\ref{fig:circuit_phase_estimation}.
It sounds so good for quantum computation
since handling many qubits
is considered to be difficult.

Here we discuss robustness of the semi-classical
implementation against the decoherence.
This implementation saves the number of qubits
dramatically, although depth of circuits
is almost the same as usual implementation.
Since influences of the decoherence is considered
to increase in proportion to product
of the depth of circuits and the number of qubits,
the semi-classical implementation is more robust
against the decoherence
from this point of view.

We next focus on the ordering of application
of controlled-$U$ operations.
On the semi-classical circuit,
controlled-$U$ operations must be applied
from controlled-$U^{2^{p-1}}$ operations
just as the descending-order case
so that measurements are done as fast as possible.
Since the semi-classical implementation
has single qubit for the first register,
most influences of the decoherence are caused
by the second register.
It means that the semi-classical circuit
is less robust against the decoherence
like the descending-order.

We finally note the robustness
of the semi-classical implementation
for the quantum counting
especially in order to solve
{\bfseries NP}-complete problems.
Checking whether the number of solutions for these problems
is zero or non-zero by the quantum counting
helps us solve the problems.
As we have shown in Proposition~\ref{thm:position_outputs_second},
the descending-order circuit has wrong peaks at
0 or $N$ independently of oracle.
Since the semi-classical circuit
is restricted to descending-order,
it may not suitable for such problems.


\section{Concluding remarks}
\label{sec:conclusion}

In this paper, we focused on
investigating
influences of the decoherence
related to the first order term of error rate
on the quantum counting
and revealing the difference of robustness
against decoherence
on two {\it equivalent} implementations.

In the analysis of decoherence on the first register,
we showed that probability distribution on the quantum counting
had wrong peaks caused by the depolarizing channel
at a distance of the power of two from correct peaks,
and the probability distribution was
independent of ordering of
application of controlled-$G$ operations.
In the analysis on the second register,
we first showed that
wrong outputs were obtained near the correct one
on the ascending-order circuit,
whereas
particular wrong outputs $0$ and $N$
were obtained with high probability
on the descending-order circuit.
We then clarified that
the correct output were obtained
with higher probability
on the ascending-order circuit
than on the descending-order one.
Consequently, the ascending-order implementation of the quantum counting
was more robust against the decoherence.

We also discussed the decoherence on the phase estimation.
Similar to the quantum counting,
the probability to estimate the desired phase
by the phase estimation
such as the factoring
was higher on the ascending-order circuit.
Moreover,
we pointed out weakness
of the semi-classical implementation against the decoherence
because of the descending-order.

\ack

We are very grateful to Prof.{}\ Hiroshi Imai
for giving helpful advice to us,
and also thank Dr.{}\ Jumpei Niwa
for providing Quantum Computation Simulation System (QCSS)
for numerical calculations to us.


\appendix


\setcounter{section}{0}


\section{Probability distribution in the first register case}
\label{sec:appendix_first}

We consider the case that
the decoherence error based on the depolarizing channel
occurs once on the first register.
Suppose that
the error occurs on the $j$-th qubit
of the first register
after application of controlled-$G^k$
with the $j$-th control qubit,
as shown in Figure~\ref{fig:position_error1st}.
Under the depolarizing channel,
$\sigma_x, \sigma_y, \sigma_z$ errors
disturb each qubit
with the same probability.
We calculate probability distributions
in each error case
and take an average of probability distributions
of all cases for simulating the depolarizing channel with pure states.

\subsection{$\sigma_x, \sigma_y, \sigma_z$ errors and identity $\sigma_0$}

We first deal with $\sigma_x$ error.
Let $\sum_{j=0}^{p-1} m_j2^j := m$
be indexes of the first register
in the quantum counting.
The quantum state before $\sigma_x$ error is represented as follows:
\begin{equation*}
 \amp{P} \sum_{m=0}^{P-1} \ket{m} \tensor G^{km_j + (m \bmod 2^j)} \ket{s}.
\end{equation*}
Since $\sigma_x$ error flips a bit $m_j$ of the quantum state $\ket{m}$,
the state is disturbed by $\sigma_x$ into:
\begin{equation*}
 \amp{P} \sum_{m=0}^{P-1} \ket{m+(1-2m_j)2^j}
  \tensor G^{ km_j + (m \bmod 2^j) } \ket{s}.
\end{equation*}
The rest of controlled-$G$ operations
and the QFT are applied to this state,
\begin{eqnarray*}
 &\eqspc \amp{P} \sum_{m=0}^{P-1} \ket{m + (1-2m_j)2^j}
 \tensor G^{ m+(1-2m_j)(2^j-k)  } \ket{s} \nonumber \\
 &\stackrel{F.T.}{\rightarrow} \sum_{m'=0}^{P-1}
 \ket{m'} \tensor
 \sum_{m=0}^{P-1}
 \sum_{l=\pm} \ceffirst{l}{x}\ket{l},
\end{eqnarray*}
where $\ket{\pm} := \amp{2}(\ket{b}\mp{}i\ket{g})$ and
$\ceffirst{\pm}{x} := \frac{1}{\sqrt{2}P}
\epimmP e^{\pm\rmi\theta{}/2} e^{\pm\rmi\theta{}(m-(1-2m_j)k)}$.

By calculation of $|\ceffirst{\pm}{x}|$, we have
\begin{eqnarray}
 \left|\ceffirst{\pm}{x}\right|
  &= \left|\frac{1}{\sqrt{2}P} \summ
      \left[ \epimmP e^{\pm\rmi\theta{}(m-(1-2m_j)k)} \right]
     \right| \nonumber \\
 &= \amp{2} \cdot
 \sinL{(m'\pm{}f)} \times \cosmj{\left(m'\pm{}f\pm{}2k\frac{f}{2^j}\right)} \nonumber \\
 &\eqspc \quad \times \sinM{(m'\pm{}f)}.
 \label{eq:x_first}
\end{eqnarray}

Similarly, we calculate $|\ceffirst{\pm}{i}|$
where $i=y,z$, corresponding to $\sigma_i$ errors.
\begin{eqnarray}
 \left|\ceffirst{\pm}{y} \right| &= \amp{2} \cdot
  \sinL{(m'\pm{}f)} \times \sinmj{\left(m'\pm{}f\pm{}2k\frac{f}{2^j}\right)} \nonumber \\
  &\eqspc \quad \times
  \sinM{(m'\pm{}f)}.
\end{eqnarray}
\begin{eqnarray}
 \left|\ceffirst{\pm}{z} \right| &= \amp{2} \cdot
 \sinL{(m'\pm{}f)} \times \cosmj{(m'\pm{}f)} \nonumber \\
 &\eqspc \quad \times \sinM{(m'\pm{}f)}.
\end{eqnarray}
We also calculate $|\ceffirst{\pm}{0}|$
for symmetry of the depolarizing channel
in Equation~(\ref{eq:depolarizing}).
\begin{eqnarray}
 \left|\ceffirst{\pm}{0}\right| &=
  \amp{2} \cdot \frac{\sin[\pi(m'+f)]}{2^p \sin[\pi (m'+f) /2^p]}
  \nonumber \\
 &= \amp{2} \cdot
 \sinL{(m'\pm{}f)} \times \sinmj{(m'\pm{}f)} \nonumber \\
 &\eqspc \quad \times \sinM{(m'\pm{}f)}.
\end{eqnarray}

\subsection{Probability distribution of all errors}

Let $Prob^{(i, j, k)}$ be the probability
to observe $m'$ as a measurement result,
where $i=x,y,z,0$, $0 \le j \le p-1$, and $0 \le k \le 2^p-1$.
Taking summation of probability distributions in all cases,
we obtain the following probability distribution
on the depolarizing channel:
\begin{eqnarray}
 \fl \frac{1}{4} \sum_{i=0, x, y, z} Prob^{(i, j, k)}(m') 
  =
  \frac{1}{4} \sum_{i=x,y,z,0}\sum_{l=\pm}
	       \left|\ceffirst{l}{i}\right|^2 \nonumber \\
 \lo= \frac{1}{4} \left[
  \sinL{(m'+f)} \times \sinM{(m'+f)} \right]^2 \nonumber\\
 +
  \frac{1}{4} \left[
  \sinL{(m'-f)} \times \sinM{(m'-f)} \right]^2.
\end{eqnarray}
$\frac{1}{4} \sum_{i=0, x, y, z} Prob^{(i, j, k)}(m')$
does not depend on $k$.
It means that this probability distribution
is uniquely determined by the qubit where the error occurs,
independently of when the error occurs.


\section{Probability distribution
 in the second register case}
\label{sec:appendix_second}

Here we consider to calculate
probability distributions of the quantum counting
when the decoherence occurs on the second register.
We deal with two probability distributions
on the ascending-order circuit and the descending-order one.

\subsection{The ascending-order circuit}
\label{sec:appendix_second_asc}

We first consider the ascending-order circuit
for the quantum counting.
Suppose that
the decoherence error occurs on the second register
after application of controlled-$G^k$ operations
with the $j$-th control qubit,
similar to the case of first register.

Let $r$ be the number of application of
controlled-$G$ operations
according to the first register $\ket{m}$
before the decoherence occurs.
\begin{equation*}
 r := km_j + (m \bmod 2^j).
\end{equation*}
The quantum state in the quantum counting before the decoherence
is represented as follows:
\begin{eqnarray*}
 \amp{P} \summ \ket{m}
 \tensor \left[
 \cos \left( r+\frac{1}{2} \right) \theta \ket{b}
 + \sin \left( r+\frac{1}{2} \right)\theta \ket{g} \right].
\end{eqnarray*}
By Lemma~\ref{lem:Grover_six_bases}, the decoherence disturbs the state
into
\begin{equation}
 \amp{P} \summ \ket{m}
 \tensor \left[
 \cos \left( r+\frac{1}{2} \right) \theta \ket{b'}
 + \sin \left( r+\frac{1}{2} \right)\theta \ket{g'} \right],
\end{equation}
where
\begin{eqnarray*}
 \ket{b'} &:= \fourGroverstate{b}, \\
 \ket{g'} &:= \fourGroverstate{g},
\end{eqnarray*}
satisfying Equation~(\ref{eq:six_properties}).
By application of
the rest of controlled-$G$ operations and the QFT,
\begin{eqnarray}
 \fl \frac{1}{P} \summp \summ \epimmP \ket{m'} \tensor
 \left[ \cos \rhalf \theta \cdot G^{m-r} \ket{b'} +
 \sin \rhalf \theta \cdot G^{m-r} \ket{g'} \right] \nonumber \\
 \lo= \summp \ket{m'} \tensor \frac{1}{2P} \summ \epimmP \times \nonumber \\
 \bigg[ e^{\frac{1}{2}\rmi\theta} \eirt \bigg\{
 \amp{2}\eimrt \left\{ (\ub+\rmi\vb)-\rmi(\ug+\rmi\vg) \right\}\ket{+}
 \nonumber \\
 \qquad \quad +
 \amp{2}\emimrt \left\{ (\ub-\rmi\vb)-i(\ug-\rmi\vg) \right\}\ket{-} \nonumber \\
 \qquad \quad +
           \epimr \left(\ueb\ketebb -\rmi\ueg\ketebg \right)
         + \left(\veb\ketegb -\rmi\veg\ketegg \right) \bigg\} \nonumber \\
 + e^{-\frac{1}{2}\rmi\theta} \emirt \bigg\{
 \amp{2}\eimrt \left\{ (\ub+\rmi\vb)+\rmi(\ug+\rmi\vg) \right\}\ket{+}
 \nonumber \\
 \qquad \quad +
 \amp{2}\emimrt \left\{ (\ub-\rmi\vb)+\rmi(\ug-\rmi\vg) \right\}\ket{-} \nonumber \\
 \qquad \quad +
           \epimr \left(\ueb\ketebb +\rmi\ueg\ketebg\right)
         + \left(\veb\ketegb +\rmi\veg\ketegg \right) \bigg\} \bigg].
 \label{eq:state_second_asc}
\end{eqnarray}
We focus on peaks of each term in this equation.

\subsubsection{Peaks of the terms $\ket{\pm}$}
\mbox{} \\
\vspace{-0.5mm} \mbox{} \\
Except global factors,
a coefficient of the term $\ket{+}$ in Equation~(\ref{eq:state_second_asc})
is as follows:
\begin{eqnarray*}
 &\eqspc \frac{\ub+\rmi\vb-\rmi(\ug+\rmi\vg)}{2\sqrt{2}P} \epifP \summ e^{\piP \{mm'+
 mf\}} \\
 &+ \frac{\ub+\rmi\vb+\rmi(\ug+\rmi\vg)}{2\sqrt{2}P} \empifP \summ
  e^{\piP \{mm' + (m-2r)f\}}.
\end{eqnarray*}
Here we calculate only each summation
instead of whole equation
because we want to know where the term $\ket{+}$ have peaks.
More precisely,
we calculate the first summation
\begin{eqnarray}
 \left|\summ \exp \left[ \piP\{m(m'+f) \} \right]\right|
  = \frac{ \sin[\pi(m'+f)] }{ 2^p\sin[\pi(m'+f)/2^p] },
 \label{eq:sec_asc_alpha_1}
\end{eqnarray}
and the second summation
\begin{eqnarray}
 \dasc{+} &:=
  \left|\summ \exp \left[ \piP \left\{ mm' + (m-2r)f \right\} \right]\right| \nonumber
  \\
 &= \sinL{(m'+f)} \times
 \cosmj{ \left\{m' + \left(1-\frac{2k}{2^j}\right)f\right\} } \nonumber \\
 &\eqspc \times \sinM{ (m'-f) }.
 \label{eq:sec_asc_alpha_2}
\end{eqnarray}
The first summation has single peak at $m'=-f \equiv 2^p-f$
and the second summation has a strong peak at $m'=f$
and weak peaks near $m'=f$,
as shown in Figure~\ref{fig:sec_asc_alpha}.
It is easily seen that
the term $\ket{-}$ also has two peaks at $m'=f, 2^p-f$
and weak peaks near $m'=2^p-f$.


\subsubsection{Peaks of the terms $\ketebb$ and $\ketebg$}
\mbox{} \\
\vspace{-0.5mm} \mbox{} \\
%
%
We then consider peaks of the term $\ketebb$
which consists of two summations
except global factors:
\begin{eqnarray}
 \frac{1}{P}\summ e^{\piP \{mm'+rf+(m-r)\frac{P}{2}\} },
 \quad
 \frac{1}{P}\summ e^{\piP \{mm'-rf+(m-r)\frac{P}{2}\} },
 \label{eq:sum_asc_eb}
\end{eqnarray}
Since the term $\ketebg$ also have two same summations,
we describe $\keteb$ as terms $\ketebb$ and $\ketebg$
for calculating peaks of the terms together.

Let $\dasc{e_b}$ be the first summation
in Equation~(\ref{eq:sum_asc_eb}).
\begin{eqnarray}
 \fl \dasc{e_b} :=
  \left|\frac{1}{P} \summ
 \exp\left[ \piP \left\{mm'+ rf + (m-r)\frac{P}{2} \right\} \right]\right|
 \nonumber \\
 \lo\equiv
 \sinL{ \left(m'+\frac{P}{2}\right) }
 \times
 \cosmj{ \left\{ m'+\frac{k}{2^j}f +
 \left(1-\frac{k}{2^j}\right)\frac{P}{2} \right\} } \nonumber \\
 \lo\times
 \sinM{ (m'+f) }.
 \label{eq:dasc_eb}
\end{eqnarray}
Figure~\ref{fig:sec_asc_eb}
shows a probability distribution of $\dasc{e_b}$,
which has a strong peak at $m'=2^p-f$
and weak peaks near the peak.
The second summation in the term $\dasc{e_b}$
also has a strong peak at $m'=f$
and weak peaks near the peak.

\subsubsection{Peaks of the terms $\ketegb$ and $\ketegg$}
\mbox{} \\
\vspace{-0.5mm} \mbox{} \\
The terms $\ketegb$ and $\ketegg$,
denoted by $\keteg$,
have two summations:
\begin{eqnarray}
 \frac{1}{P}\summ e^{\piP \{mm'+rf\} },
 \quad
 \frac{1}{P}\summ e^{\piP \{mm'-rf\} }.
 \label{eq:sum_asc_eg}
\end{eqnarray}
By calculation of summations,
we obtain two strong peaks at $m'=f,2^p-f$
and weak peaks corresponding to the peaks.

\subsubsection{Peaks of all terms}
\label{sec:asc_all}
\mbox{} \\
\vspace{-0.5mm} \mbox{} \\
In the case of the ascending-order circuit,
all six terms in Equation~(\ref{eq:state_second_asc})
have only two strong peaks at
$m'=f, 2^p-f$ that are the same peaks
in no error case
and weak peaks near two peaks.
Hence the overall probability distribution
of the quantum state
in Equation~(\ref{eq:state_second_asc})
is considered to have two peaks
at $m'=f, 2^p-f$.

\subsection{The descending-order circuit}

We then investigate probability distributions
in the descending-order case.
Let $r'$ be the number of controlled-$G$ operations
according to the first register $\ket{m}$
before the error.
\begin{equation*}
 r' := m - 2^jm_j - (m \bmod 2^j) + km_j.
\end{equation*}
By simple calculation,
we obtain the similar final state
in Equation~(\ref{eq:state_second_asc}),
which uses $r'$ instead of $r$.
We consider peaks of four terms
$\ketebb, \ketebg, \ketegb$, and $\ketegg$ in the equation
because the terms $\ket{\pm}$
differ little from the case of the ascending-order.

\subsubsection{Peaks of the terms $\ketebb$ and $\ketebg$}
\mbox{} \\
\vspace{-0.5mm} \mbox{} \\
Like the ascending-order circuit,
the term $\keteb$ has two summations in this case
by replacing $r$ by $r'$ in Equation~(\ref{eq:dasc_eb}).
Let $\ddes{e_b}$ be the first summation of the term.
\begin{eqnarray}
 \fl \dasc{e_b} := \left|\frac{1}{P} \summ
 \exp\left[ \piP \left\{mm'+ r'f + (m-r')\frac{P}{2} \right\} \right] \right|
 \nonumber \\
  \lo= \sinL{ (m'+f) }
 \times
 \cosmj{ \left\{ m'+\frac{k}{2^j}f + \left(1-\frac{k}{2^j} \right)
 \frac{P}{2} \right\} } \nonumber \\
 \times \sinM{ \left(m'+\frac{P}{2}\right) }.
 \label{eq:ddes_eb}
\end{eqnarray}
$\dasc{e_b}$ has a strong peak at $m' = -P/2\equiv2^p/2$,
as shown in Figure~\ref{fig:sec_des_eb}.
This peak corresponds to $-1$
in the Grover operator
in Equation~(\ref{eq:G_four_bases}).
The second summation also has the same peak at $2^p/2$.

\subsubsection{Peaks of the terms $\ketegb$ and $\ketegg$}
\begin{figure}
 \begin{minipage}{0.49\textwidth}
  \begin{center}
   \hspace{-3mm}
   \scalebox{1.1}{\includegraphics[width=\textwidth]{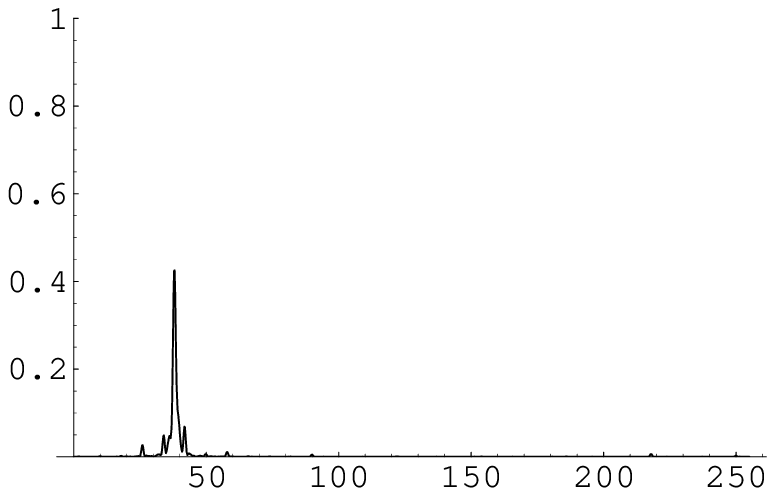}}

   \vspace{-0.5mm}
   $m'$
   \vspace{-1mm}
  \end{center}
  \caption{The probability distribution of
  $\dasc{+}$.}
  \label{fig:sec_asc_alpha}
 \end{minipage}
 \begin{minipage}{0.49\textwidth}
  \begin{center}
   \scalebox{1.1}{\includegraphics[width=\textwidth]{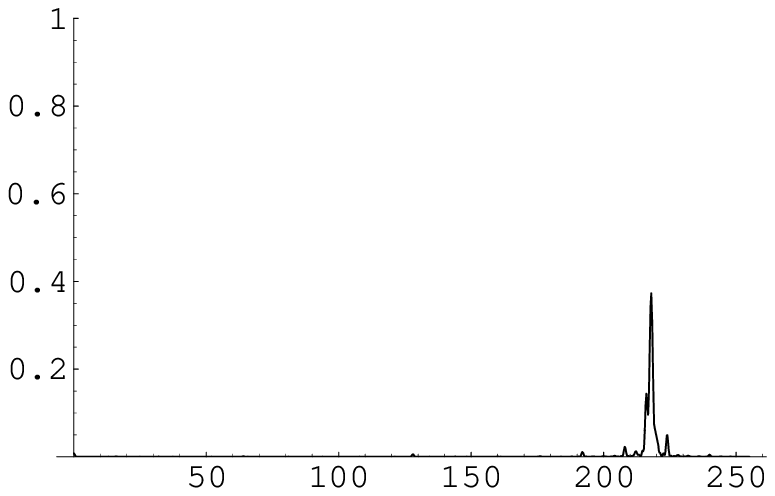}}

   \vspace{-0.5mm}
   \hspace{4mm}$m'$
   \vspace{-1mm}
  \end{center}
  \caption{The probability distribution of
  $\dasc{e_b}$.}
  \label{fig:sec_asc_eb}
 \end{minipage}

 \vspace{4mm}
 \begin{minipage}{0.49\textwidth}
  \begin{center}
   \hspace{-3mm}
   \scalebox{1.1}{\includegraphics[width=\textwidth]{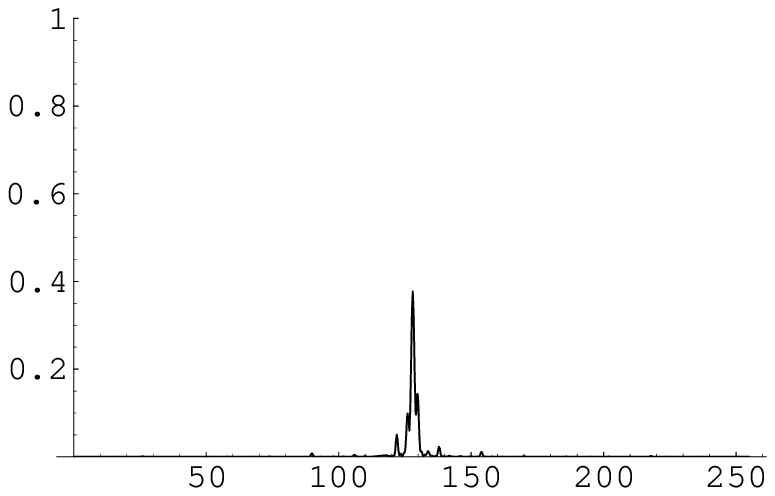}}

   \vspace{-0.5mm}
   $m'$
   \vspace{-1mm}
  \end{center}
  \caption{The probability distribution of
  $\ddes{e_b}$.}
  \label{fig:sec_des_eb}
 \end{minipage}
 \begin{minipage}{0.49\textwidth}
  \begin{center}
   \scalebox{1.1}{\includegraphics[width=\textwidth]{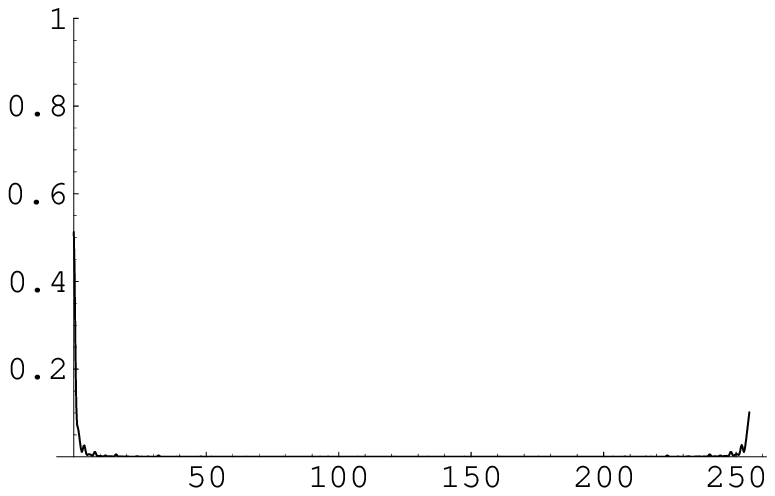}}

   \vspace{-0.5mm}
   \hspace{4mm}$m'$
   \vspace{-1mm}
  \end{center}
  \caption{The probability distribution of
  $\ddes{e_g}$.}
  \label{fig:sec_des_eg}
 \end{minipage}
\end{figure}

\mbox{} \\
\vspace{-0.5mm} \mbox{} \\
Let $\ddes{e_g}$ be a summation of the term $\keteg$
corresponding to $\dasc{e_g}$ in Equation~(\ref{eq:sum_asc_eg}).
\begin{eqnarray}
 \ddes{e_g} &= \sinL{ \left(m'+\frac{P}{2}\right) }
  \times \cosmj{ \left( m'+\frac{k}{2^j}f \right) } \nonumber \\ 
 &\eqspc \times \sinM{ m' }.
\end{eqnarray}
We show a probability distribution of $\ddes{e_g}$
in Figure~\ref{fig:sec_des_eg},
and obtain a peak at $m'=0$.
The second summation in $\keteg$
also has the same peak at $m'=0$.


\subsubsection{Peaks of all terms}
\mbox{} \\
\vspace{-0.5mm} \mbox{} \\
The overall probability distribution in the case of the descending-order
has four strong peaks at $m'= f, 2^p-f$, i.e.{}\ the correct peaks,
and $m'= 0, 2^p/2$, particular wrong peaks
that do not appear in the ascending-order case and the first register
case.
The positions of these wrong peaks are fixed
independently of an oracle and type of errors.

\pagebreak

\vspace{1mm}
\bibliographystyle{plain}


\begin{thebibliography}{10}

\bibitem{AW99integrals}
D.~S. Abrams and C.~P. Williams.
\newblock Fast quantum algorithms for numerical integrals and stochastic
  processes.
\newblock arXiv:quant-ph/9908083, 1999.

\bibitem{azuma02decoherence}
H.~Azuma.
\newblock Decoherence in {G}rover's quantum algorithm: Perturbative approach.
\newblock {\em Phys. Rev. A}, \textbf{65}(4, 042311), 2002.

\bibitem{Beauregard022nplus3}
S.~Beauregard.
\newblock Circuit for {Shor's} algorithm using 2n+3 qubits, 2002.

\bibitem{BHT98counting}
G.~Brassard, P.~H{\o}yer, and A.~Tapp.
\newblock Quantum counting.
\newblock In {\em Proceedings of the 25th International Colloquium on Automata,
  Languages and Programming, Lecture Notes in Computer Science}, volume
  \textbf{1443}, pages 820--831, 1998.

\bibitem{DH96minimum}
C.~D{\"{u}}rr and P.~H{\o}yer.
\newblock A quantum algorithm for finding the minimum.
\newblock arXiv:quant-ph/9607014, 1996.

\bibitem{Grover96grover}
L.~K. Grover.
\newblock A fast quantum mechanical algorithm for database search.
\newblock In {\em Proceedings of the 28th Annual ACM Symposium on Theory of
  Computing}, pages 212--219, 1996.

\bibitem{Grover96median}
L.~K. Grover.
\newblock A fast quantum mechanical algorithm for estimating the median.
\newblock arXiv:quant-ph/9607024, 1996.

\bibitem{Heinrich02summation}
S.~Heinrich.
\newblock Quantum summation with an application to integration.
\newblock {\em Journal of Complexity}, \textbf{18}:1--50, 2002.

\bibitem{NC00computation}
M.~A. Nielsen and I.~L. Chuang.
\newblock {\em Quantum Computation and Quantum Information}.
\newblock Cambridge University press, 2000.

\bibitem{NMI02simulator}
J.~Niwa, K.~Matsumoto, and H.~Imai.
\newblock General-purpose parallel simulator for quantum computing.
\newblock {\em Phys. Rev. A}, \textbf{66}, 062317, 2002.

\bibitem{Niwa02ErrorCorrect}
J.~Niwa, K.~Matsumoto, and H.~Imai.
\newblock Simulating the effects of quantum error-correction schemes.
\newblock arXiv:quant-ph/0211071, 2002.

\bibitem{OD99simulating}
K.~M. Obenland and A.~M. Despain.
\newblock Simulating the effect of decoherence and inaccuracies on a quantum
  computer.
\newblock {\em Lecture Notes in Computer Science}, 1509:447--459, 1999.

\bibitem{PP00efficient_factorization}
S.~Parker and M.~B. Plenio.
\newblock Efficient factorization with a single pure qubit and $log {N}$ mixed
  qubits.
\newblock {\em Phys. Rev. Lett.}, \textbf{14}:3049--3052, 2000.

\bibitem{SMB03noise}
D.~Shapira, S.~Mozes, and O.~Biham.
\newblock Effect of unitary noise on {G}rover's quantum search algorithm.
\newblock {\em Phys. Rev. A}, \textbf{67}(4, 042301), 2003.

\bibitem{Shor94factorization}
P.~W. Shor.
\newblock Algorithms for quantum computation: Discrete logarithms and
  factoring.
\newblock In {\em Proceedings of the 35th {IEEE} Symposium on Foundations of
  Computer Science}, pages 124--134, 1994.

\bibitem{SZL98decoherence}
C.P. Sun, H.~Zhan, and X.F. Liu.
\newblock Decoherence and relevant universality in quantum algorithms via a
  dynamic theory for quantum measurement.
\newblock {\em Phys. Rev. A}, \textbf{58}(3), 1998.

\bibitem{TW02path}
J.~F. Traub and H.~Wo{\'{z}}niakowski.
\newblock Path integration on a quantum computer.
\newblock arXiv:quant-ph/0109113, 2002.

\bibitem{YS99SU}
S.~Yu and C.P. Sun.
\newblock Quantum searching's underlying {SU(2)} structure and its quantum
  decoherence effects.
\newblock arXiv:quant-ph/9903075, 1998.

\end{thebibliography}

\end{document}